\documentclass[pdflatex,sn-basic]{sn-jnl}

\usepackage{graphicx}%
\usepackage{multirow}%
\usepackage{amsmath,amssymb,amsfonts}%
\usepackage{amsthm}%
\usepackage{mathrsfs}%
\usepackage[title]{appendix}%
\usepackage{xcolor}%
\usepackage{textcomp}%
\usepackage{booktabs}%
\usepackage{algorithm}%
\usepackage{algorithmicx}%
\usepackage{algpseudocode}%
\usepackage{listings}%
\usepackage{url}
\usepackage{hyperref}

\begin{document}

\title[Article Title]{Far-right party influence on polarization dynamics in electoral campaign}

\author*[1,2]{Eva Rifà}\email{eva.rifa@eurecat.org}

\author[1,2]{Joan Massachs}\email{joan.massachs@est.fib.upc.edu}

\author[2,3]{Emanuele Cozzo}\email{emanuele.cozzo@ub.edu}

\author[1]{Julian Vicens}\email{julian.vicens@eurecat.org}

\affil*[1]{Eurecat, Centre Tecnològic de Catalunya, Barcelona, Spain}

\affil[2]{Universitat de Barcelona Institute of Complex Systems (UBICS), Barcelona, Spain}

\affil[3]{CNSC-IN3, Universitat Oberta de Catalunya, Barcelona, Spain}


\abstract{Political polarization has attracted increasing attention in recent years, driven by the rise of social media and the global emergence of far-right populist movements. This study investigates the dynamics of structural polarization during electoral campaigns in multi-party systems, with a particular focus on the presence of far-right actors and their influence on polarization patterns and hate speech. Using retweet networks as a measure of structural polarization, we analyze two case studies in Spain: the 2022 Andalusia regional elections, where the far-right party Vox was a significant contender, and the 2019 Barcelona city council elections, where the party had no representation. Our results reveal that the presence of a far-right party intensifies polarization, leading to the formation of two distinct ideological blocks aligned along left-right ideological axes, as observed in Andalusia. In contrast, the Catalan independence movement in Barcelona diluted the alignment of voters, resulting in a more complex, multi-axis polarization landscape. We also explore the relationship between polarization and hate speech, finding an anti-correlation between them in both cases. Our findings underscore the significant role of far-right movements in driving political polarization and the nuanced effects of different political contexts on polarization dynamics.}


\maketitle

\section{Introduction}

The study of political and ideological polarization has garnered significant attention from both academics and the general public since the early 1990s~\citep{c061d6ea-7871-3144-85cf-f5a32ab78d79}. This interest has been particularly intense in the United States, where polarization has been closely associated with the more general phenomena of Culture Wars, that have shaped political discourse over the past few decades. Some scholars argue that the end of the Cold War disrupted the broad consensus that had previously underpinned political stability, contributing to the rise in polarization together with the rise of the religious right~\citep{du2023jesus}. In recent years, the focus on this issue has been revitalized by the global emergence of populist far-right movements, such as Donald Trump's election as U.S. President in 2016, and by the widespread adoption of social media.

While social media platforms provide an invaluable lens through which to observe political and ideological polarization, there is no scientific consensus on whether social media use itself has a polarising effect~\citep{acerbi2019cultural}. Several studies, including \cite{del2016echo}, have reported empirical evidence suggesting that platforms like Facebook and Twitter contribute to increased polarization. However, other research such as that of \cite{tucker2018social}, has either failed to find such an effect or has even pointed in the opposite direction, indicating that the relationship between social media use and polarization is complex and context and topic-dependent~\citep{barbera2015tweeting}. Nevertheless, it is widely recognised that political campaigns tend to exacerbate polarization, regardless of the medium through which they are conducted~\citep{hansen2017campaigns}.

Traditionally, measures of polarization have focused on the collective, or distributional, properties of individual attitudes within a society~\citep{c061d6ea-7871-3144-85cf-f5a32ab78d79}. Specifically, in the context of political or ideological polarization, these measures examine the distributional properties of public opinion. Among the various operationalisations proposed, the family of measures presented by \cite{esteban1994measurement} define polarization as reaching its maximum when a population is divided into two equally sized groups positioned at the greatest possible ideological distance from each other. With the advent of social big data, a range of new measures has emerged to capture polarization dynamics. One such measure is structural polarization, which assesses the extent to which a network of endorsements, such as retweets, splits into two distinct and opposing groups~\citep{polarization}.  It is also well-established that the retweet mechanism serves as a broadcasting tool, indicating that users who retweet not only endorse the content of the original tweet but also show a significant level of interest, actively opting to share the information with their followers~\citep{MARTINGUTIERREZ2023113244}. This perspective is common in the Twitter analysis literature, where retweets are viewed as evidence of a user supporting the message of the original poster~\citep{falkenberg_growing_2022}. For this reasons, retweets networks are well suited to study polarization. 

In two-party systems, such as that of the United States, the increase in polarization during electoral campaigns can be seen as obvious or even trivial, driven by the binary nature of political competition~\citep{olivares2019opinion}. However, in multi-party systems, the emergence and nature of polarization are less straightforward~\citep{hansen2017campaigns} and are related to the relation between in-parties and out-parties~\citep{Martin-Gutierrez2024-zc}. Electoral campaigns in these contexts can result in a range of outcomes, from block polarization, where parties align into two opposing blocs, to fragmentation or multipolarization~\citep{MARTINGUTIERREZ2023113244}, where voter attention becomes more evenly dispersed across multiple parties. The predominance of research focused on U.S. data, however, has often led to an overlooking of these complexities, leaving a gap in understanding how polarization manifests in multi-party systems electoral campaigns.

Thus, it is particularly important to study electoral campaigns in political systems where new actors have emerged, especially populist far-right movements. These actors often disrupt traditional party dynamics and can significantly influence patterns of polarization. Furthermore, polarization may or may not be accompanied by an increase in hate speech, making it of interest to explore the relationship between the two. Those frequently employ identity based divisive rhetoric in opposition to emerging political actors previously unrepresented in the political sphere, like feminists, racialised people, and migrants~\citep{una2019leia,nagle2017kill}. This can include framing issues in ways that appeal to specific ideological or partisan groups, heightening emotional responses, and demonising opponents~\citep{romero2023process}. That could exacerbate both polarization and the prevalence of hate speech in political discourse~\citep{bjornsgaard_media_nodate}.

In this study, we will investigate polarization during electoral campaigns in Spain, that experience the presence of new parties, including a populist far-right one, alongside the other traditional parties. We examine the impact of the presence of a far-right party on polarization from the point of view of the dynamical retweet network. On one hand, we will study the case of the 2022 regional elections in Andalusia, being specially relevant because they were the first elections after the entrance of the far-right party Vox in the parliament of Andalusia. On the other hand, we will examine the case of the 2019 Barcelona City Council elections, where the far-right party had no representation in the city council. Also, we study the relationship between hate speech and polarization during these periods. For simplicity, we will refer to the Andalusia 2022 regional elections dataset to simply Andalusia and for the Barcelona 2019 City Council elections to Barcelona.

\subsection{Andalusia 2022 Regional Elections}

The 2022 Andalusia Regional Election was held on Sunday, 19 June. These elections select the members of the Parliament of Andalusia, which in turn determines the composition of the \textit{Junta de Andalucía}, the institution that organizes the self-government of the Autonomous Community of Andalusia, in Spain.

In the 2018 elections, for the first time in 36 years, right-wing parties secured a majority, leading to the end of the long-standing dominance of the \textit{Partido Socialista Obrero Español de Andalucía} (PSOE–A). Despite being the most popular party in the region, PSOE-A was replaced in government by a coalition led by Juan Manuel Moreno of the \textit{Partido Popular de Andalucía} (PP). Moreno assumed the presidency, forming an alliance with \textit{Ciudadanos} (Cs), and relying on support from Vox for a confidence and supply arrangement. The main parties and coalitions participating in the 2022 elections, along with their ideologies, were as follows: PSOE-A, social democracy; PP, liberal conservatism; Cs, liberalism and autonomism; \textit{Por Andalucía} (PorA), left-wing populism and green politics; \textit{Adelante Andalucía} (AA), Andalusian nationalism, left-wing populism, and anti-capitalism; and Vox, right-wing populism, ultranationalism, and national conservatism (Section S2).

In the 2022 elections, the PP won for the first time in all electoral constituencies and also achieved its first absolute majority in the autonomous community. The PSOE-A got its worst result ever in the autonomous community, while Vox failed to fulfil expectations and saw only modest gains. Support for Cs collapsed, with the party being left out of parliament, whereas the left-wing vote divided between PorA and AA platforms~\citep{enwiki:1229380969, enwiki:1229380959}. 

\subsection{Barcelona 2019 City Council Elections}
The 2019 Barcelona City Council Election was held on Sunday, 26 May. This election followed the Spanish general elections held a month earlier, which marked the first time Vox secured seats in the Spanish parliament. In the previous City Council elections of 2015 \textit{Barcelona en Comú} (BComú), led by Ada Colau, emerged as the dominant party.

The main candidacies for 2019 elections were: \textit{Esquerra Republicana de Catalunya} (ERC), catalan independence, left-wing nationalism and social democracy; BComú, left-wing populism and participatory democracy; \textit{Partit dels Socialistes de Catalunya} (PSC), socioliberalism; \textit{Barcelona pel Canvi-Ciutadans} (BCNcanvi-Cs), liberal conservatism and Spanish unionism; \textit{Junts per Catalunya} (Junts), catalan independence liberalism; \textit{Partit Popular de Catalunya} (PP), liberal conservatism; \textit{Capgirem BCN-AMUNT} (CUP), catalan independence anti-capitalism and socialism and \textit{Barcelona és Capital-Primàries} (BCAP), catalan independence (Section S3). 

Under the leadership of Ernest Maragall, ERC achieved a historic victory in the 2019 elections as it was the first time it had won since the Franco dictatorship. This shift came as the resurgent PSC drew significant support away from the incumbent Ada Colau and her BComú party. Despite this, Colau managed to retain the mayorship by forming an alliance with PSC and securing support from BCNcanvi councillors, led by Manuel Valls. Valls, who had previously run for the French presidency in 2017, was nominated by Cs as their mayoral candidate. Valls's support of Colau's investiture was based on his stated intention to prevent the pro-Catalan independence camp from securing control over Catalonia's capital city~\citep{enwiki:1229560773, enwiki:1229560925}.
 
\section{Methods}
\subsection{Data}

The data was collected using academic access to the Twitter API. For Andalusia dataset we use Twitter discussions about the 2022 regional elections held in 19 June. The data includes tweets published between 1 April and 31 July 2022, thus including one month after the elections and two months before them, that includes all the electoral campaign. The tweets included are the ones that contain any of the specific keywords considered, that are general keywords like ``19J"; party specific keywords like ``AndalucíaLiberal", one of the campaign lemmas; as well as the main list heads Twitter accounts of the majority parties that are: \textit{@JuanMa\_Moreno}, \textit{@\_JuanEspadas}, \textit{@Macarena\_Olona}, \textit{@InmaNietoC}, \textit{@TeresaRodr\_} and \textit{@JuanMarin\_}. In total, there are 8.1 million tweets.

For Barcelona dataset we followed the same collection structure as mentioned before. This includes Twitter discussions about 2019 Barcelona City Council elections held in 26 May recollected from 1 April to 30 June 2019. The tweets included are the ones containing different types of keywords, the ones considered are: general keywords like ``Municipals2019Bcn" or ``DebatTV3CatRadio"; party specific keywords like ``ERCBcn", ``BcnEnComu" or ``PSCBarcelona"; and the user account of the heads of the main political parties: \textit{@ErnestMaragall}, \textit{@AdaColau}, \textit{@JaumeCollboni}, \textit{@ManuelValls}, \textit{@QuimForn}, \textit{@JosepBouVila}, \textit{@AnnaSaliente}, \textit{@JordiGraupera} and \textit{@IGarrigaVaz}. In total, there are 3.3 million tweets.

\subsection{Communities}
We use a community detection algorithm to identify groups of users from the retweets network. Our goal is to obtain communities that describe the political leanings of network users. To do it, we take the largest weakly connected component and use the Clauset-Newman-Moore greedy modularity maximisation~\citep{clauset_finding_2004} to find a community partition. This algorithm begins with each node in its community and repeatedly joins the pair of communities that lead to the largest modularity until no local-scale increase in modularity is possible.

The use of modularity maximisation for community detection has been the subject of significant debate, as highlighted by \cite{peixoto_tiago_2021}, as it does not fully account for the statistical implications of the optimisation process in identifying the best partition. Nevertheless, it proves valuable in identifying the political alignment of Twitter users based on political party organisations.

\subsection{Polarization}

Structural polarization assesses the extent to which the network of retweets splits into two distinct and opposing groups. A network is considered polarised if it consists of two communities that have many internal interactions but few interactions between them.

Hence, the same algorithm used for community detection is used to compute polarization but now with a fixed number of clusters of two. We compute the \textit{normalized} (standardised and denoised) polarization to reduce biases related to the network's characteristic features~\citep{polarization}. This polarization is calculated by standardising and denoising the modularity on partitions of random graphs generated with the same configuration as the original network. The value is computed:

\begin{equation}
    \hat{\Phi}_z(G) = \frac{\Phi(G)- \langle \Phi(G_{CM})\rangle}{\sqrt{\langle \Phi(G_{CM})^2\rangle-\langle \Phi(G_{CM})\rangle^2}}
\end{equation}

We compute the change point of the polarization and hate speech percentage time series with the Python package ruptures~\citep{truong_selective_2020}. For Andalusia we have manually set the penalty value to 5 in order to visualise the trend changes due to the election date (June 19, 2022). For Barcelona, the penalty value has been set to 2 for the same reason, which is the election date (May 26, 2019).

\subsection{Hate Speech Detection}

We define a measure of the amount of hate in the network as the percentage of tweets that contain words that are identified as hate. To compute if a tweet contains hate we use a multilingual BERT model~\citep{mbert} that was trained to classify text inputs as containing hate speech or not with datasets of 5365 and 6000 tweets in Spanish~\citep{basile-etal-2019-semeval, s19214654}\footnote{The best validation score achieved by the algorithm trained at different learning rates is 0.740287 for a learning rate of $3 \times 10^{-5}$. More details about the model can be found at the official model repository: \url{https://huggingface.co/Hate-speech-CNERG/dehatebert-mono-spanish}.}. This method is subject to several limitations, including the fact that a significant portion of the Barcelona dataset contains tweets in Catalan without the model being trained in this language. However, we know that the algorithm is able to distinguish hate tweets in this language.

\section{Results}\label{sec2}

\subsection{Community analysis}

\begin{figure}[ht]
\centering
\begin{minipage}{.4\textwidth}
    \centering
    \includegraphics[width=\textwidth]{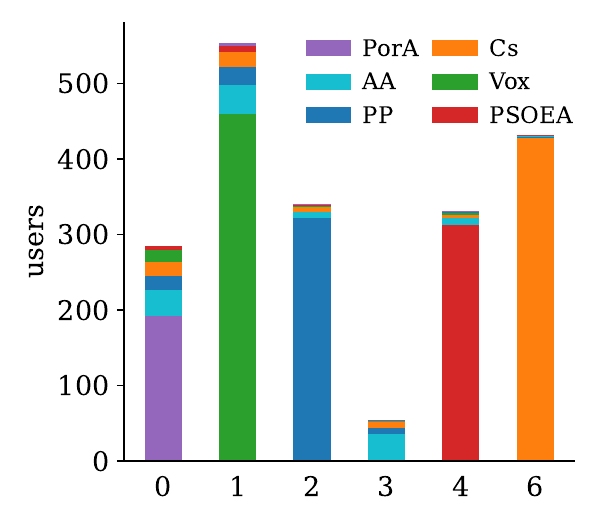}
    {\small \textbf{(a)} Andalusia}
\end{minipage} 
\begin{minipage}{.4\textwidth}
    \centering
    \includegraphics[width=\textwidth]{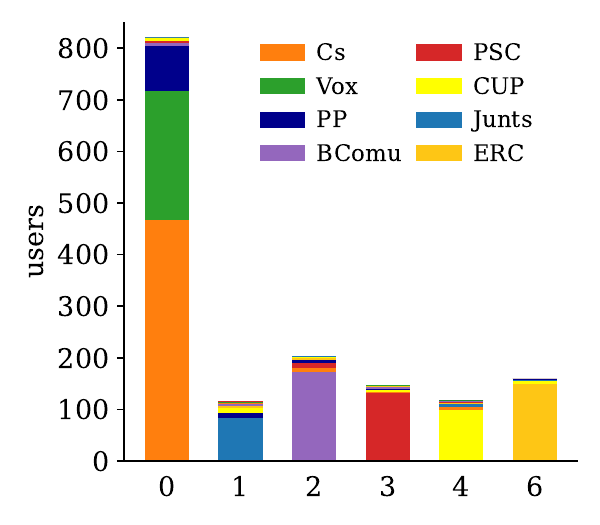}
    \small{\textbf{(b)} Barcelona}
\end{minipage}
    
\caption{Number of usernames containing the main political party names for each community.}
\label{counts_community_index}
\end{figure}

After applying the community detection algorithm, we can identify the political alignment of the network groups. This step is done assuming that the retweet is an endorsement of the idea contained in the tweet, and then we name each community with the corresponding political alignment of the accounts it contains. The naming process has been done by analysing the maximum number of user accounts that contain the name of the main political parties competing in the elections. Thus, we named each community with the name of the political party most times represented by the name of the user accounts (Fig. \ref{counts_community_index}).

\begin{figure}[ht]
\centering
\includegraphics[width=0.7\textwidth]{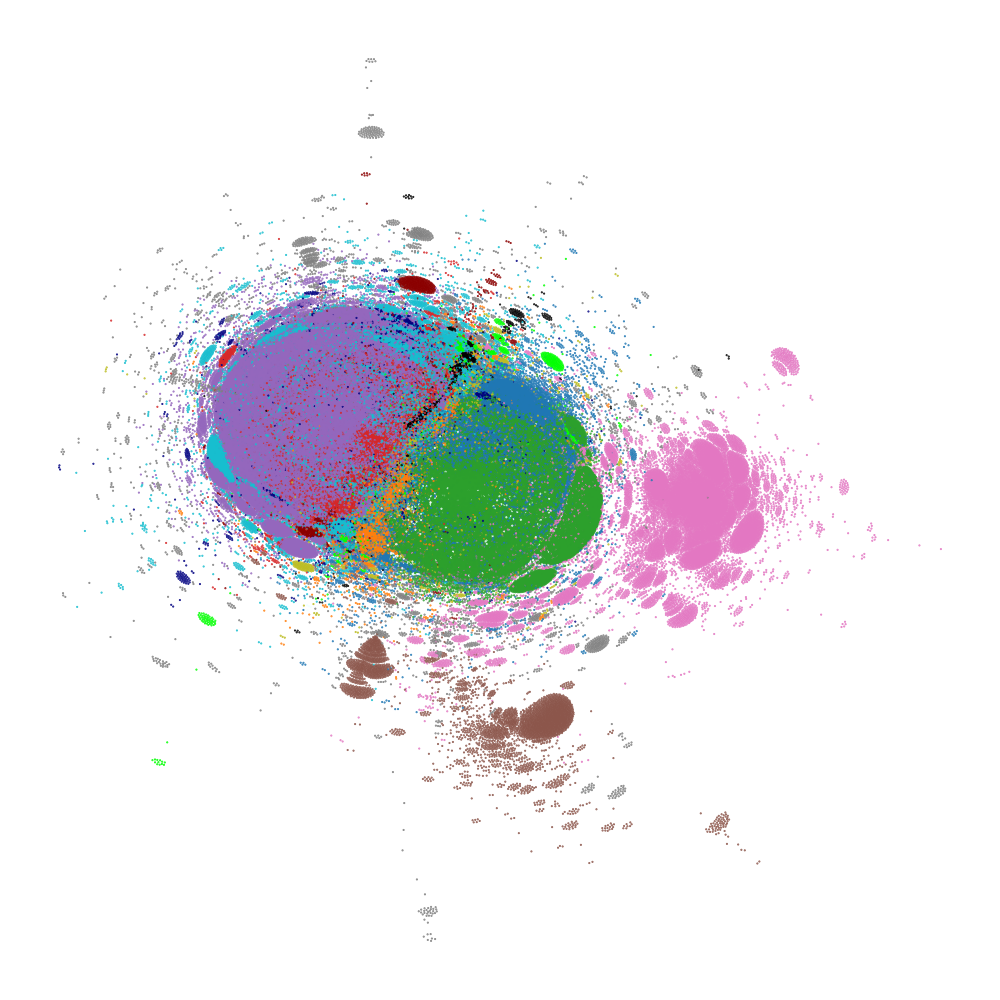}
\caption{Network of users of the Twitter discussion for Andalusia elections. The green community correspond to Vox; blue and orange communities correspond to PP and Cs; purple, red and cyan correspond to PorA, PSOE-A and AA.}
\label{communities-andalucia}
\end{figure}

\begin{figure}[ht]
\centering
\includegraphics[width=0.7\textwidth]{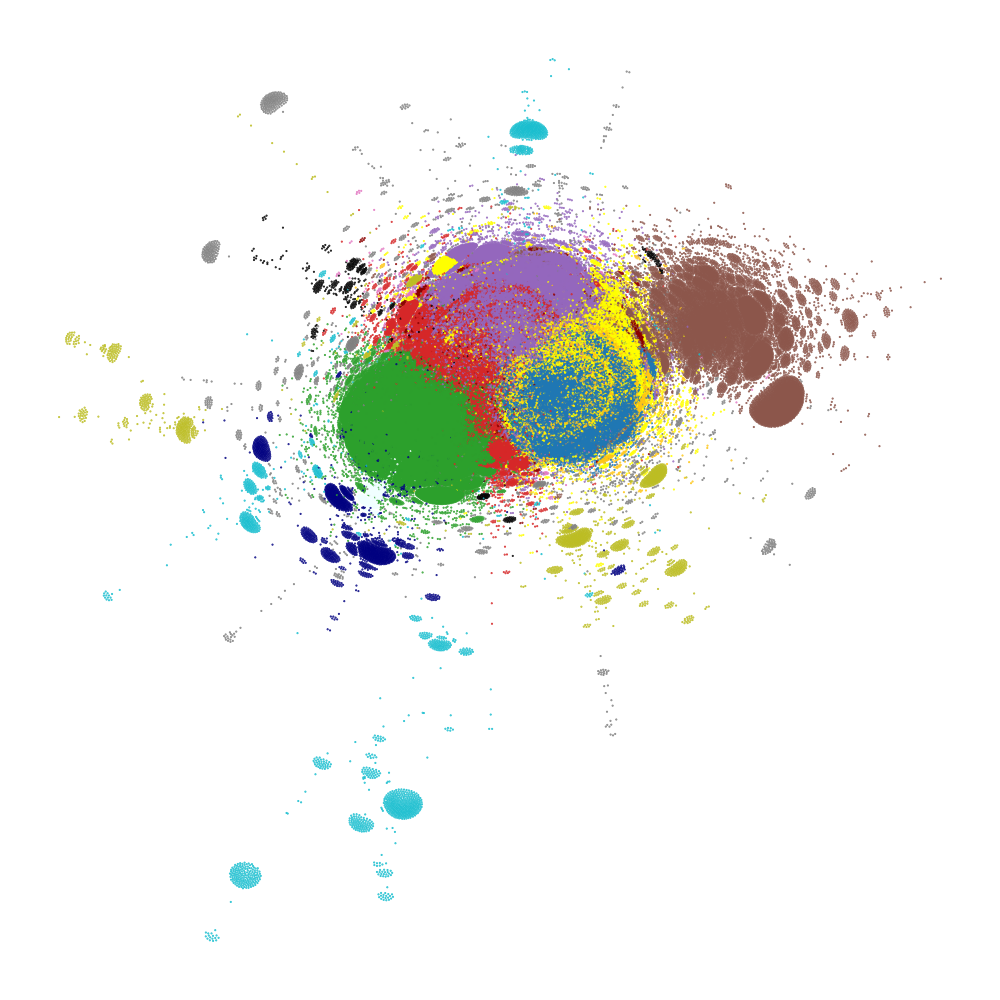}
\caption{Network of users of the Twitter discussion for Barcelona elections. The green community corresponds to Cs, blue are from Junts, purple corresponds to BComú, red corresponds to PSC, yellow corresponds to CUP and amber corresponds to ERC.}
\label{communities-bcn}
\end{figure}

In Figs. \ref{communities-andalucia} and \ref{communities-bcn} we present the network of users and their communities for Andalusia and Barcelona, respectively. Each node represents a user, and they are connected if one has retweeted a tweet from the other. The weight of the edge is the number of retweets. The layout is the SFDP spring-block layout, and the colour of the user corresponds to the community identified by a modularity maximisation algorithm. More detailed information about the communities and their political alignments is provided in Tables \ref{table-centers-andalucia} and \ref{table-centers-bcn}.

\begin{table}[ht]
\centering
 \begin{tabular}{c l l l l l l} 
 \hline
 i & Colour & Name & Primary node & Secondary node & Users & Retweets \\ [0.5ex] 
 \hline
 1 & green & Vox & Macarena$\_$Olona & AndaluciaVox & 89236 & 1529059 \\ 
 0 & purple & PorA & InmaNietoC & JA$\_$DelgadoRamos & 113617 & 453149 \\ 
 4 & red & PSOE-A & AndaluciaSinVOX & $\_$JuanEspadas & 18954 & 364775 \\
 2 & blue & PP & JuanMa$\_$Moreno & ppandaluz & 24480 & 138929  \\
 6 & orange & Cs & CiudadanosCs & JuanMarin$\_$ & 6516 & 47107 \\ 
 3 & cyan & AA & TeresaRodr$\_$ & AdelanteAND & 22013 & 44408 \\
 8 & olive &  & OpositaJA & Crisstinabj & 2979 & 20197 \\ 
 9 & navy &  & RecortesCero & LaEtxebarria & 2792 & 17299 \\ 
 5 & pink &  & PizarroMariaJo & vanedelatorre & 15656 & 14798 \\
 12 & lime &  & Gatajusticie & JaenMereceMas & 1344 & 2624 \\ 
 13 & azure &  & santi544 & ASLETKUBIS & 1276 & 2623 \\
 11 & black &  & unidadporc1 & Sonia18586892 & 1394 & 1588 \\ 
 10 & darkred &  & PartidoPACMA & DaniNovarama & 1502 & 1398 \\ 
 7 & brown &  & cubacooperaveCO & cubacooperaveCJ & 3999 & 1315 \\  [1ex]

 \hline
 \end{tabular}
 \caption{Description of communities for Andalusia dataset, ordered by retweet count. The network is constructed with retweets as edges, including only communities with more than 1000 nodes. The first column indicates the order of communities by number of users.}
 
 \label{table-centers-andalucia}
\end{table}

For  Andalusia the largest community in users is PorA, followed by Vox, PP, AA, PSOE-A and Cs, in order. This list includes all the main parties in the electoral campaign, each represented in different communities. Other smaller communities have not been clearly associated with any political party or organisation and are set with no label in Table \ref{table-centers-andalucia}. Looking at the layout, one can observe a main cluster of communities with a central division between the left and right. In the centre-right, there is Vox (green), alongside PP. Next to them, at the centre-left, there is PorA (purple) surrounded by AA (cyan). The communities located in the center include PSOE-A (red) in the left division and Cs (orange) on the right division of the main cluster.

For Barcelona, the largest community by number of users consists of a combination of Cs, PP and Vox, all of which are right-wing Spanish nationalist parties. For simplicity, we refer to this community as Cs, as it has the most representation among the three in both the dataset and the election results. Similarly, the Junts community also includes the BCAP candidacy. The parties are ranked in order as follows: Junts, BComú, Vox, ERC, CUP and PSC. The main right-wing parties are clustered next to PSC. Additionally, there is a vertical separation between BComú and the pro-independence parties at the bottom of the layout. It is interesting to note that closeness between different communities strongly depends on their views regarding Catalonia's independence. While in Fig. \ref{communities-andalucia} we see PSOE-A and PorA (Podemos) together opposing Vox, here PSC and BComú are separated. Furthermore, we observe that Junts and CUP are positioned in the same region of the layout, despite being ideologically distinct parties: CUP is anti-capitalist, while Junts is right-wing. The right-wing Spanish nationalist parties, along with representatives from Vox in the centre, form the largest community by number of users. There is a large unidentified community in brown corresponding to French accounts, which we believe is not part of the electoral process in Barcelona.

\begin{table}[ht]

\centering
 \begin{tabular}{c l l l l l l} 
 \hline
 i & Colour & Name & Primary node & Secondary node & Users & Retweets \\ [0.5ex] 
 \hline
 1 & blue & Junts & JordiGraupera & jordiborras & 50152 & 1331293 \\ 
 0 & green & Cs &  Igarrigavaz & GuajeSalvaje & 52029 & 482532 \\
 2 & purple & BComú & AdaColau & bcnencomu & 42399 & 329330 \\ 
 3 & red & PSC & pscbarcelona & jaumecollboni & 23442 & 156472 \\
 6 & amber & ERC & ernestmaragall & ERCbcn & 12384 & 131053 \\
 4 & yellow & CUP & HiginiaRoig & CUPBarcelona & 19089 & 129807\\
 5 & brown &  & Furtif  & ccastanette & 15396 & 25316 \\
 10 & darkred &  & nurygglez &  greenpeace$\_$esp & 1264 & 5061 \\
 7 & olive &  & BrentToderian & voxdotcom & 2015 & 2531 \\
 9 & navy &  & TITORODRIGUEZZ & Greenpeace & 1908 & 2489 \\ 
 8 & cyan &  & DeadlineDayLive & alex$\_$orlowski & 1991 & 2253 \\  [1ex] 
 \hline
 \end{tabular}
 \caption{Political description of community centres for Barcelona dataset ordered by retweets number.}
 \label{table-centers-bcn}
\end{table}

\begin{figure}[ht]
\begin{minipage}{.5\textwidth}
    \centering
    \includegraphics[width=\linewidth]{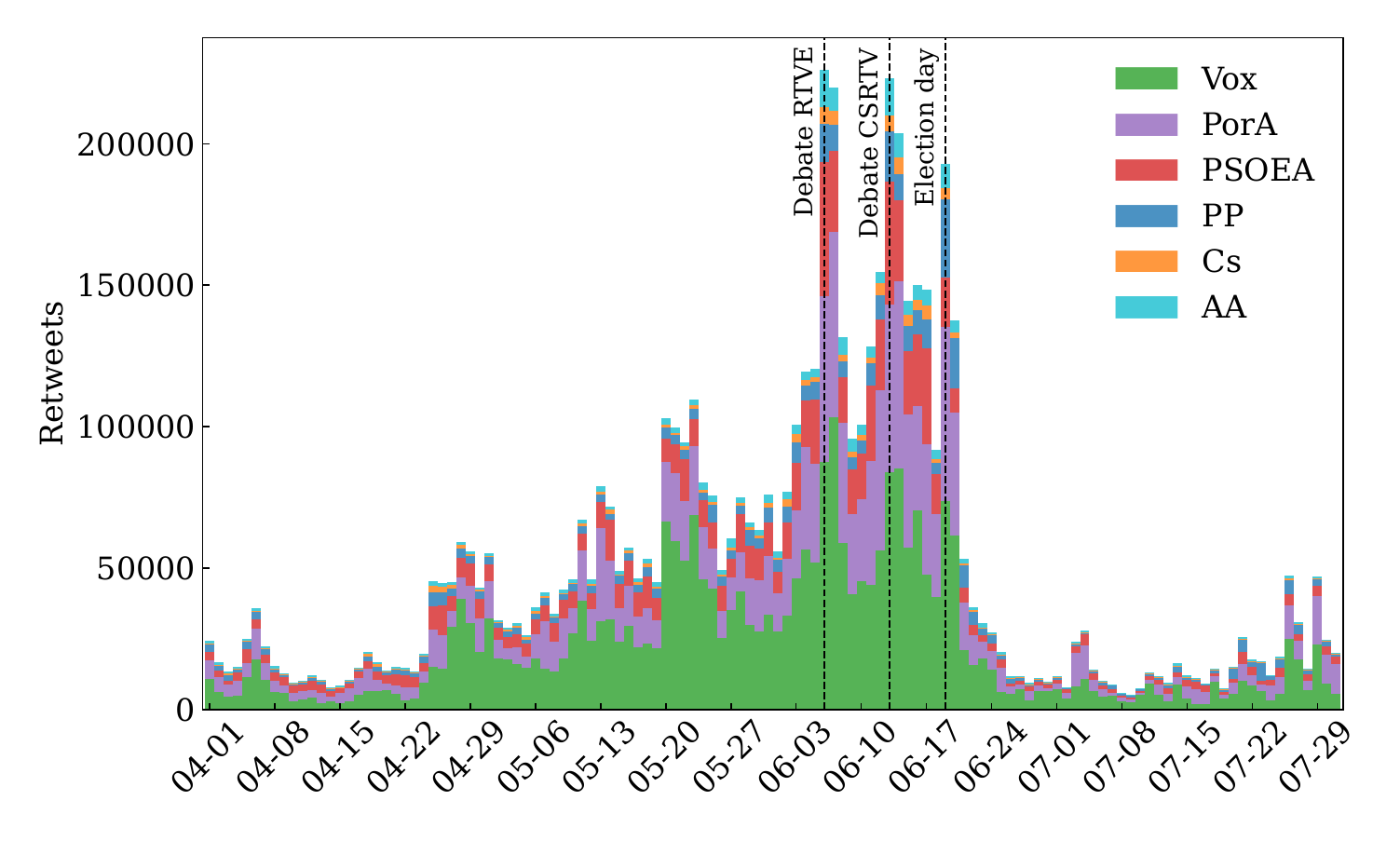}
    \small{\textbf{(a)} Andalusia}
\end{minipage}
\hfill    
\begin{minipage}{.5\textwidth}
    \centering
    \includegraphics[width=\linewidth]{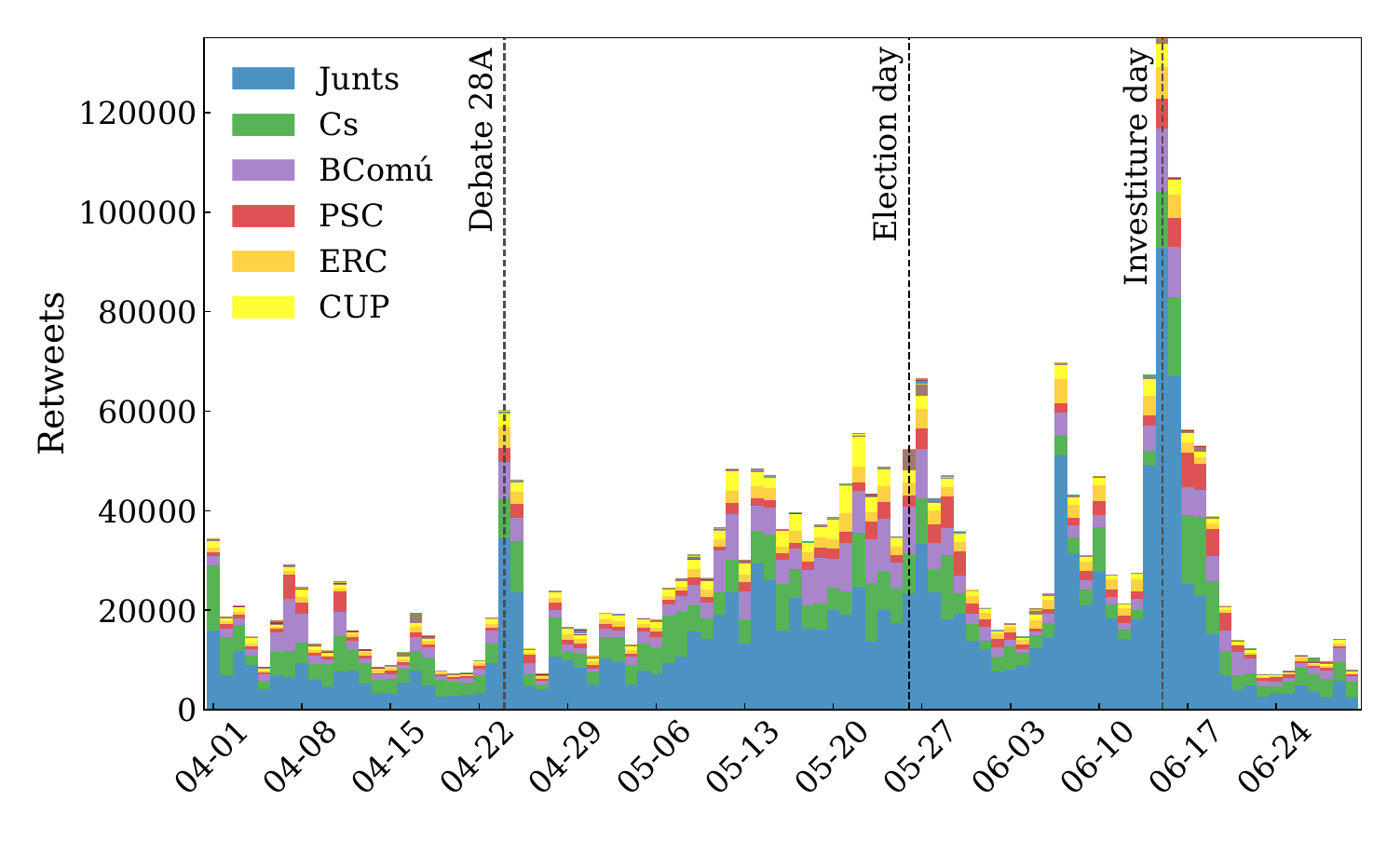}
    \small{\textbf{(b)} Barcelona}
    \end{minipage}
    
    \caption{Number of retweets per day during the electoral period by community.}
    \label{timeseries_fraction_daily}
\end{figure}

Finally, we compute the daily fraction of retweets made by each community as shown in Fig. \ref{timeseries_fraction_daily}. For Andalusia we can see that the most relevant actors throughout the entire period were Vox and PorA, with higher activity levels prior to the elections compared to after it. In Barcelona, the most active communities during all the period are Junts followed by Cs. Unlike in Andalusia, the network remained active after election day, reaching its highest peak on June 15, coinciding with the investiture day. On that day, the Junts community was the most active.

\subsection{Polarization and network size}

\begin{figure}[ht]
\begin{minipage}{.5\textwidth}
    \centering
    \includegraphics[width=\linewidth]{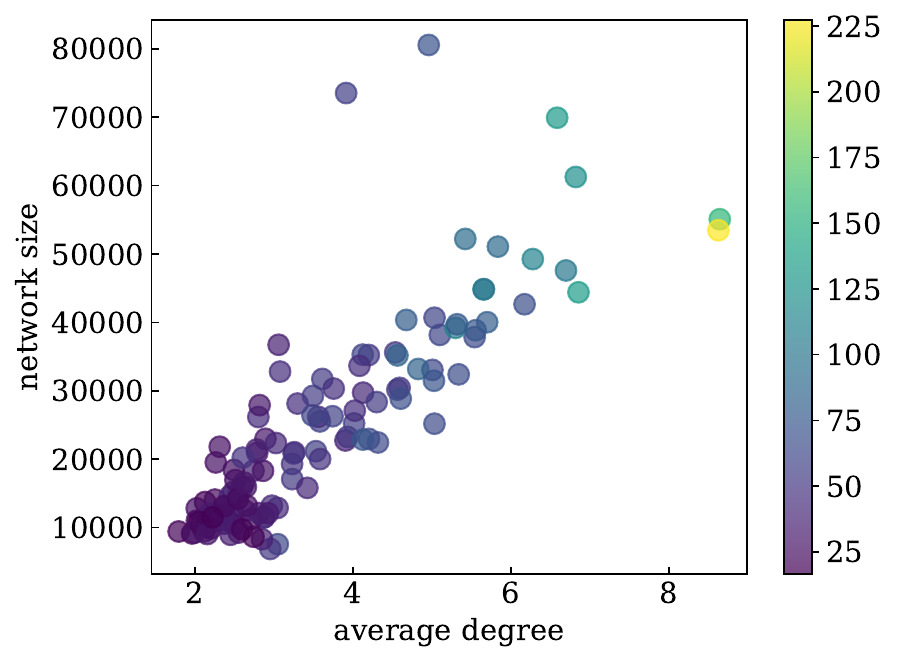}
    \small{\textbf{(a)} Andalusia}
\end{minipage}
\hfill    
\begin{minipage}{.5\textwidth}
    \centering
    \includegraphics[width=\linewidth]{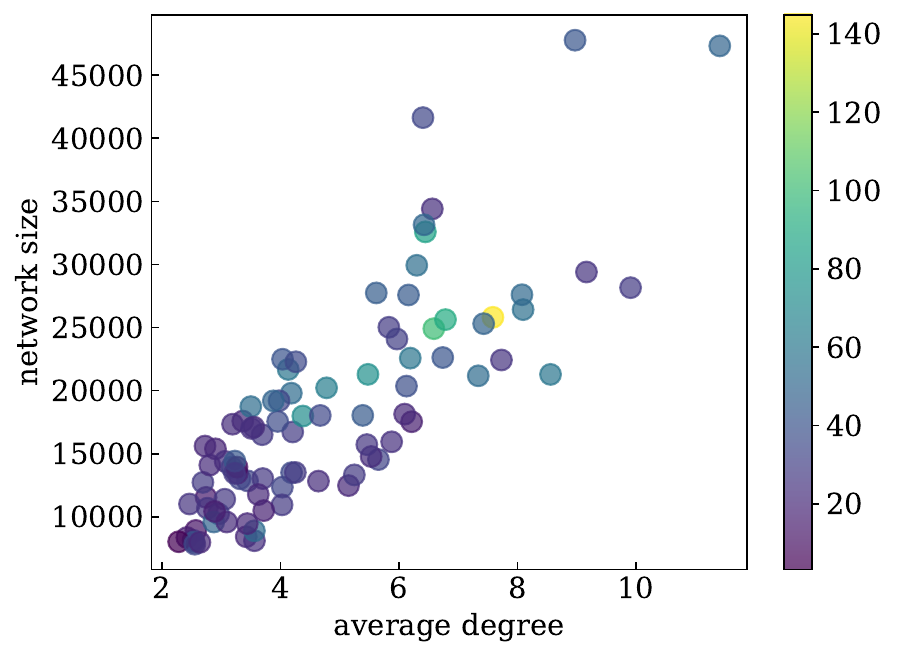}
    \small{\textbf{(b)} Barcelona}
\end{minipage}
    \caption{The structural polarization for each day as a function of average total degree and network size.}
    \label{size_and_pol}
\end{figure}

Our first question is whether the polarization depends on the network size. That is if polarization increases as more users enter the conversation. We are interested on the evolution of polarization across the studied period and for this we divide the data in days, obtaining a network with the Twitter discussion data for each day. To investigate if polarization is correlated with network features we plot the polarization as a function of network size and average degree in Fig. \ref{size_and_pol}.

For Andalusia dataset, network size is correlated with polarization. Bigger networks have a higher polarization score than smaller networks, with a Pearson Correlation Coefficient (PCC) of 0.76 and a p-value of $6 \times 10^{-24}$. The correlation of the average degree of the network with the polarization is 0.92, with a p-value of $3 \times 10^{-50}$ (see Table \ref{tab:pol_network_size_degree}, where values in parentheses represent p-values, and statistically significant results (p $<$ 0.01) are highlighted in bold).

For Barcelona, the correlation between the network size and polarization is 0.44 with a p-value of $1 \times 10^{-5}$. The average degree correlates with polarization with a PCC value of 0.43, with a p-value of $2 \times 10^{-5}$. Here, the largest networks are not the most polarised in the Barcelona dataset.

\renewcommand{\arraystretch}{1.2} 
\begin{table}[ht]
    \centering
    \begin{tabular}{l l l} 
    \hline
    Region & Network size & Average degree \\  [0.5ex] 
    \hline
    Andalusia & \textbf{0.7573} ($6 \times 10^{-24}$) & \textbf{0.9187} ($3 \times 10^{-50}$) \\  [0.5ex] 
    Barcelona & \textbf{0.4409} ($1 \times 10^{-5}$) & \textbf{0.4334} ($2 \times 10^{-5}$) \\  [0.5ex] 
    \hline
    \end{tabular}
    \caption{PCC of network size and average degree with polarization.}
    \label{tab:pol_network_size_degree}
\end{table}

\subsection{Polarization and hate speech}

The presence of hate speech in our society poses a serious threat to harmonious coexistence between different social groups~\citep{hate-elections}. This type of rhetoric can become especially prominent during electoral campaigns or times of heightened political polarization, where it may be exploited for partisan gains~\citep{romero2023process}.

\begin{figure}[ht]
\begin{minipage}{.5\textwidth}
    \centering
    \includegraphics[width=\linewidth]{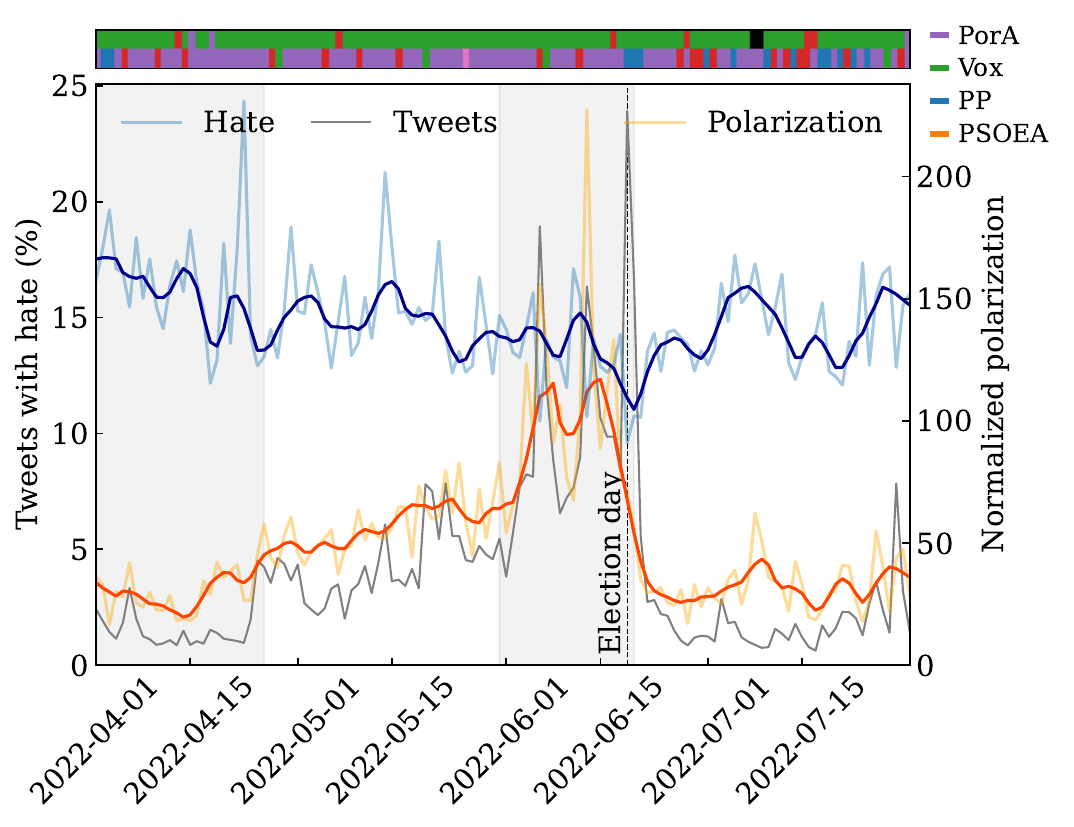}
    \small{\textbf{(a)} Andalusia}
\end{minipage}   
\begin{minipage}{.5\textwidth}
    \centering
    \includegraphics[width=\linewidth]{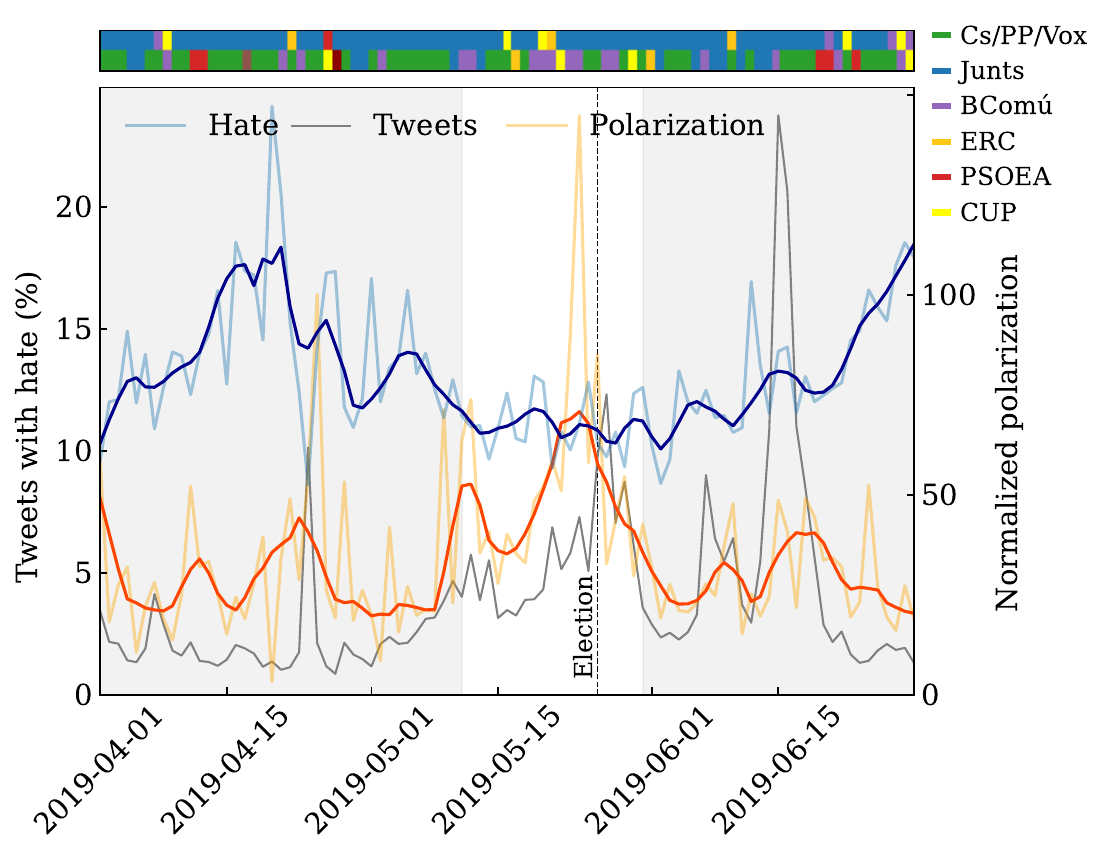}
    \small{\textbf{(b)} Barcelona}
\end{minipage}
    \caption{Comparison between the percentage of tweets with hate speech and the polarization of the interaction network for each day. The thin grey line represents the volume of tweets per day. The grey shadow mark the change points boundaries. The dashed line indicates the election date. Smoothed using Locally Weighted Scatterplot Smoothing (LOWESS) with a window of 7 days. At the top there are the communities where the daily polarization poles belong.}
    \label{fig-comparison-and-timeseries}
\end{figure}

We examine the correlation between hate speech and polarization by calculating the percentage of tweets containing hate speech and the normalized polarization for each day across both datasets, as shown in Fig. \ref{fig-comparison-and-timeseries}. 

For Andalusia, we can see that the polarization increases with the size of the network until the election day and abruptly decreasing after it. That confirms the expected impact of electoral campaigns polarising the political behaviours offline. The number of tweets containing hate speech also increases, but the number of tweets without hate speech rises even more, resulting in a decreasing percentage of tweets that contain hate speech. After the elections, we observe an increase in the percentage of hate speech, while the total number of tweets reaches a minimum and remains approximately constant. Before the elections, during the electoral campaign period, the polarization and hate speech are anti-correlated (see Table \ref{tab:hate_pol_before_after}, where the same criteria as before for significance apply). There are two peaks of polarization occurring one and two weeks before the election, coinciding with the two days of televised debates between the candidates in the Andalusian 2022 regional elections that we are analysing~\citep{rtvees_primer_2022, canalsur_debate_2022}. Close to these two peaks of polarization, we also observe peaks in the number of tweets; however no peaks of hate are observed, only small fluctuations. 

\begin{table}[ht]
    \centering
    \begin{tabular}{l l l} 
    \hline
    Region & Before elections & After elections \\  [0.5ex] 
    \hline
    Andalusia & \textbf{-0.4897} ($4 \times 10^{-6}$) & 0.0106 (0.9) \\  [0.5ex] 
    Barcelona & \textbf{-0.3805} (0.004) & -0.2509 (0.1) \\  [0.5ex] 
    \hline
    \end{tabular}
    \caption{PCC of polarization and hate before and after elections.}
    \label{tab:hate_pol_before_after}
\end{table}

In Barcelona, we observe that the percentage of hate speech is approximately constant throughout the studied period. However, there is a peak two months before the elections (April 24, 2019) coinciding with the date of the electoral debate for another election, the Spanish elections of April 28, 2019. This debate was broadcast on Catalan Public Television (TV3) and had a significant impact, becoming the most viewed general election debate in TV3's history. It was also the most commented on television broadcast of the day on Twitter across Spain~\citep{tv3_debat_2019}. Regarding polarization, we note that it reaches its highest value on the election date (June 26, 2019). Contrary to the case of Andalusia, polarization does not increase with the system size and fluctuates throughout all the entire period. Moreover, polarization increased after the elections, coinciding with the day of the investiture, which was marked by significant controversy. On this day, the number of tweets reaches its maximum, but not the level of polarization. On average, polarization in Andalusia is higher than in Barcelona, with an average value of $50 \pm 3$ compared to $35 \pm 2$. Before the elections, there is anti-correlation between hate speech and polarization.


We compute the change point of the polarization and hate speech. For Andalusia, we identify three change points: the first occurs at the beginning of the Twitter discussion, the second one month before the election date, and the third the day of the elections. We then calculate the correlation between the percentage of hate speech and polarization for each period. However, these results are not statistically significant (see Table \ref{tab:combined_breakpoints}, where the same criteria apply). In the case of Barcelona dataset, we find two change points: one month before the election date and another just after the elections. Again, these results are not statistically significant.

\begin{table}[ht]
    \centering
    \begin{tabular}{l l l l l l}
    \hline
    & \multicolumn{2}{c}{Andalusia} & \multicolumn{2}{c}{Barcelona} \\
   \hline
    Period & End date & PCC & End Date & PCC \\
    \hline
    1 & 2022-04-26 & -0.3464 (0.09) & 2019-05-11 & -0.2816 (0.08) \\
    2 & 2022-05-31 & -0.0700 (0.7) & 2019-05-31 & -0.0609 (0.8) \\
    3 & 2022-06-20 & -0.3862 (0.09) & 2019-06-30 & -0.0619 (0.7) \\
    4 & 2022-07-31 & 0.0106 (0.9) & & \\
    \hline
    \end{tabular}
    \caption{Breakpoints for timeseries of hate and polarization.}
    \label{tab:combined_breakpoints}
\end{table}

\begin{figure}[ht]
\begin{minipage}{.5\textwidth}
    \centering
    \includegraphics[width=\linewidth]{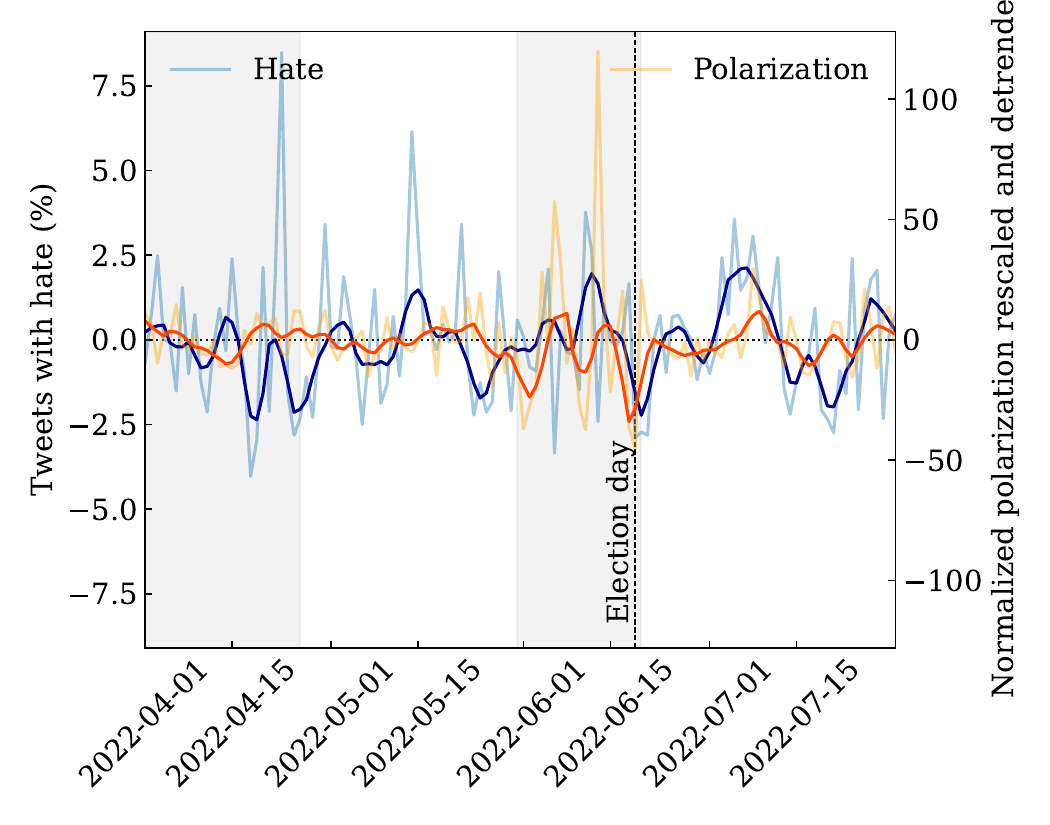}
    \small{\textbf{(a)} Andalusia}
\end{minipage}
\hfill    
\begin{minipage}{.5\textwidth}
    \centering
    \includegraphics[width=\linewidth]{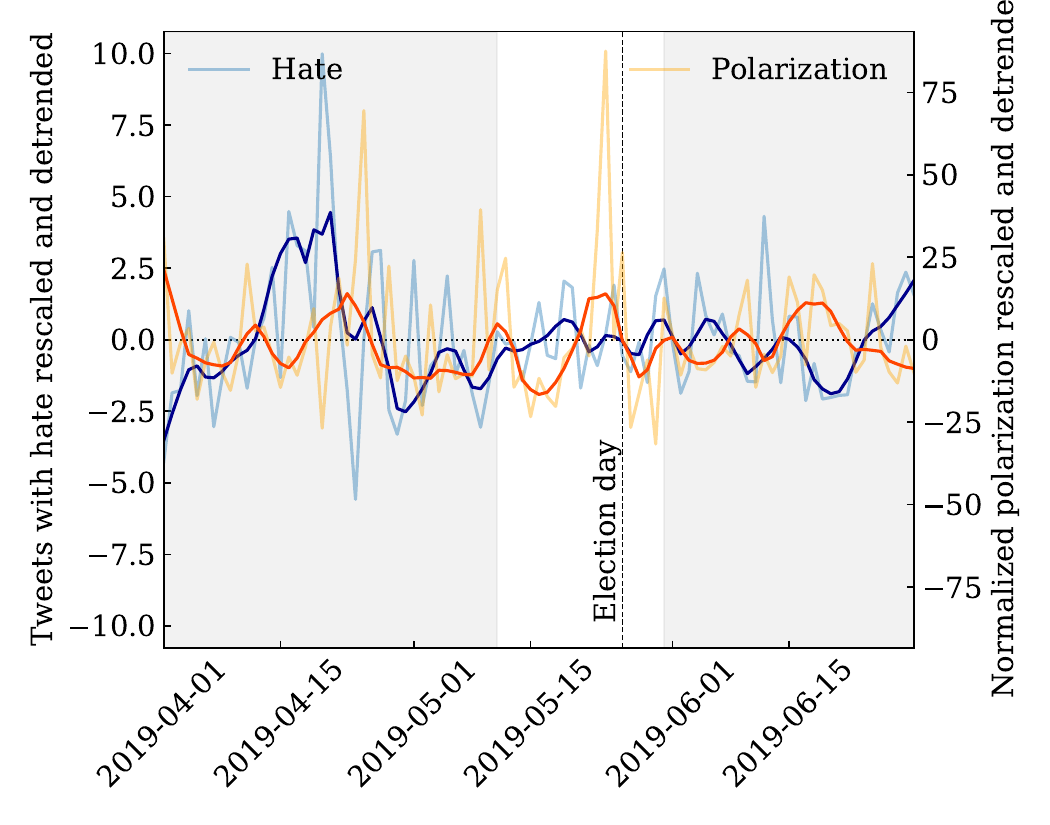}
    \small{\textbf{(b)} Barcelona}
\end{minipage}
    \caption{Comparison between the detrended and reescaled percentage of tweets with hate speech and polarization of the interaction network for each day. The grey shadow represents the period between change points.}
    \label{fig-comparison-and-timeseries-normalized}
\end{figure}

In Fig. \ref{fig-comparison-and-timeseries-normalized}, we present the detrended plot of polarization and hate speech, which results from removing the linear trend from the original time series data. This plot shows the fluctuations around the mean without the influence of the trend. The anti-correlation of polarization and hate speech is more pronounced for specific peaks in Barcelona than in Andalusia. However, the correlation between hate speech and polarization for the detrended data is not statistically significant (see Table \ref{tab:hate_pol_total_detrended}).

\begin{table}[ht]
    \centering
    \begin{tabular}{l l l} 
    \hline
    Region & Without detrending & Detrended data \\  [0.5ex] 
    \hline
    Andalusia & \textbf{-0.3113} (0.0005) & -0.1456 (0.1) \\  [0.5ex] 
    Barcelona & \textbf{-0.3306} (0.001) & -0.1817 (0.08) \\  [0.5ex] 
    \hline
    \end{tabular}
    \caption{PCC of polarization and hate for total period data with and without detrending.}
    \label{tab:hate_pol_total_detrended}
\end{table}

\subsection{Stability of communities}
\label{sec:stability} 

\begin{figure}[ht]
\begin{minipage}{.5\textwidth}
    \centering
    \includegraphics[width=\linewidth]{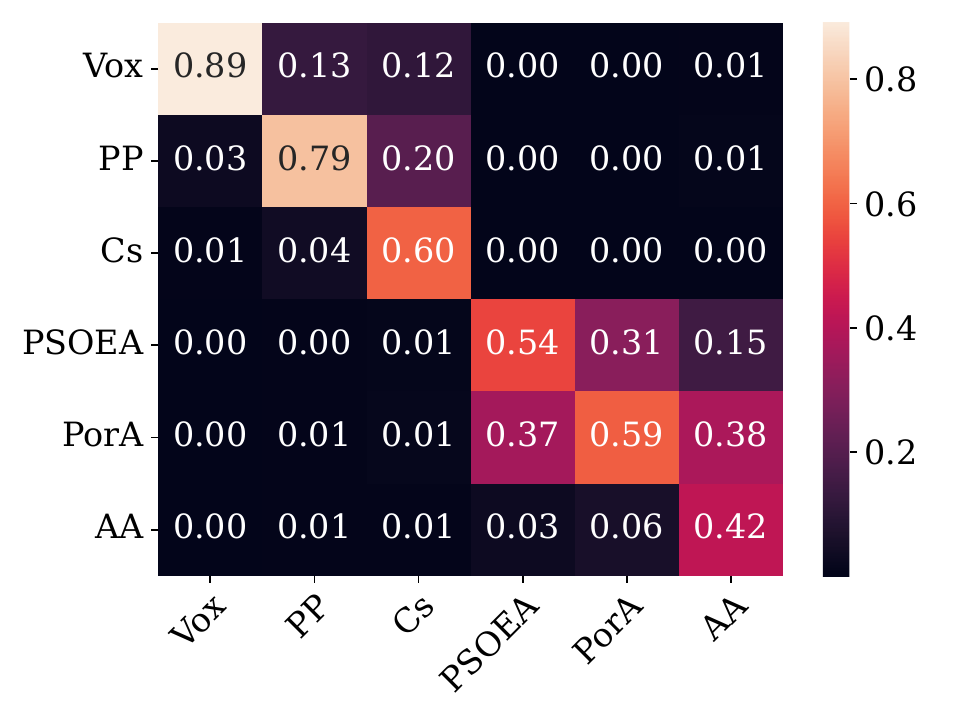}
    \small{\textbf{(a)} Andalusia}
\end{minipage}
\hfill    
\begin{minipage}{.5\textwidth}
    \centering
    \includegraphics[width=\linewidth]{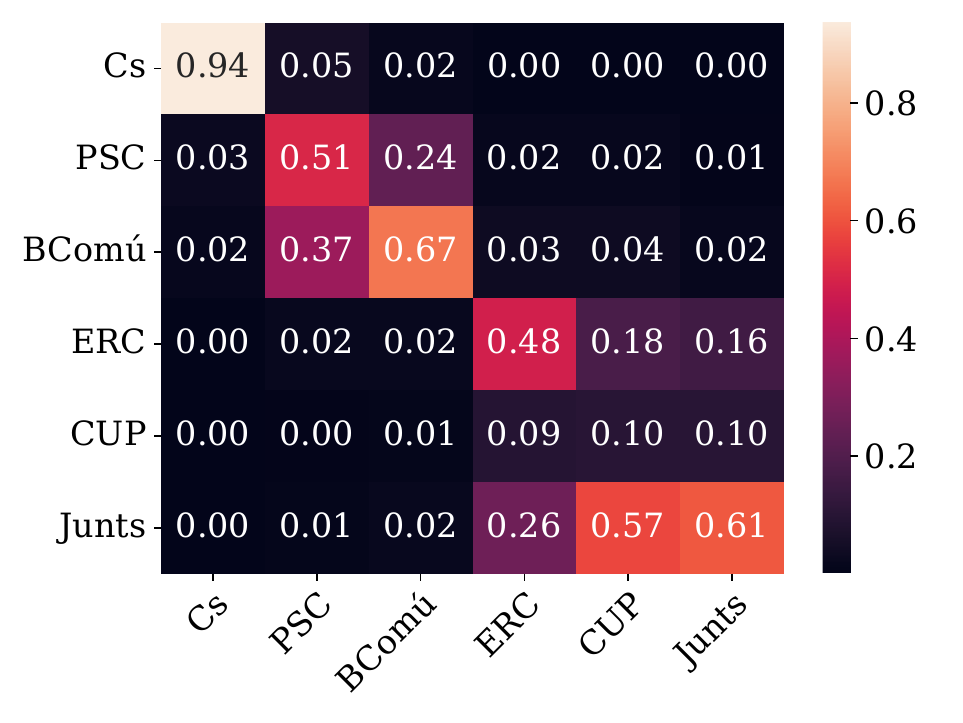}
    \small{\textbf{(b)} Barcelona}
    \end{minipage}
    \caption{Estimation of probability of a user belonging to community X in a day to become a member of the community Y the day after.}
    \label{prob-community-change}
\end{figure}

Now we calculate the probability that a user changes community from one day to the next. We can identify fluxes between communities and its stability in Fig. \ref{prob-community-change}. An extended version of this figure with all the communities detected by the algorithm can be found in the Supplementary Material (Figs. S1.1 and S1.2).
For Andalusia we see that the interchange of users happens especially inside left-wing parties: PorA and PSOE-A. There is a probability that AA users move to PorA, but not the opposite. Also, there is a probability that PP and Cs communities users move to Vox. However, there is little probability that a Vox community user move to another community. 

For Barcelona, we can see a high probability of moving from community to another for users from BComú and PSC. There is also some interchange between (in order) CUP to Junts, ERC to Junts, CUP to ERC, Junts to ERC and Junts to CUP. Here, Cs community is the most stable by difference, with almost no probability to exchange users.

\section{Discussion}
This research advances our understanding of political polarization during electoral campaigns in multi-party systems, providing insights into how these dynamics unfold, particularly in the presence of new political actors. Our findings reveal significant differences in polarization patterns between elections with and without the presence of a far-right party with a realistic chance of winning.

In the case of Andalusia, where the far-right party Vox was a significant contender, we observed a clear growing trend of structural polarization throughout the electoral campaign, peaking on election day and followed by a rapid drop. The observed pattern aligns with the trend of user engagement in the political discussion. This result is in line with other studies on polarization in dichotomous political campaigns, that used different methodologies to measure polarization~\citep{olivares2019opinion}.

Moreover, our analysis of the community structure of the retweet network highlights a pronounced two-block arrangement, with users clustering around right-wing versus left-wing ideologies. This indicates that, despite the presence of multiple parties, the election dynamics in Andalusia exhibited characteristics of block polarization~\citep{hansen2017campaigns}, effectively rendering the campaign dichotomous. The existence of these ideological blocks is further corroborated by the analysis of party change probabilities for users from day to day, which indicates a higher likelihood of shifts occurring within the same ideological group, reinforcing the two-block structure. This tendency is higher for users in the left-wing block. 

The picture is different in the case of the Barcelona city council elections. Unlike in Andalusia, where the left-right polarization axis dominates, the political landscape in Barcelona is significantly perturbed by the pro-/anti-independence axis. This dual axis of polarization reshapes the retweet network structure.

In the case of Barcelona, Spanish right-wing anti-independence parties form a unique community, which is connected to the rest of the network primarily through the left-wing anti-independence party, PSC. Interestingly, the radical left pro-independence party CUP is positioned close to the centre-right pro-independence party Junts, indicating a convergence around the independence issue. The left-wing populist and moderately anti-independence party BComú is situated near both CUP and PSC, highlighting its intermediary role between the pro- and anti-independence camps.

Further analysis of party change probabilities reveals a three-block structure: the Spanish right-wing, left-wing anti-independence, and pro-independence groups. Political polarization is generally considered higher when positions on different issues align closely with party choice~\citep{c061d6ea-7871-3144-85cf-f5a32ab78d79}. In this case, however, the presence of multiple intersecting issues dilutes the alignment, leading to lower overall polarization. Structural polarization metrics capture this complexity, with average polarization in Andalusia being notably higher compared to the lower levels observed in Barcelona.

The temporal evolution of polarization in Barcelona reveals that the dynamics are heavily influenced by specific events external to the campaign, such as the Spanish general election debate. On these key days, polarization spikes significantly, disrupting any clear overall trend of increasing polarization throughout the campaign. Consequently, in Barcelona, the largest networks do not correspond to the most polarized moments. This contrasts with the more consistent growth of polarization observed in Andalusia, where the campaign's progression steadily intensified polarization up until election day.

These findings suggest that the presence of a far-right party with a realistic chance of winning significantly polarizes the electoral discussion, driving it into two distinct ideological blocks. In contrast, a dichotomous issue like independence does not lead to the same level of heightened polarization, as other important political stances, which are transversal, remain relevant and prevent a complete alignment of the electorate into just two opposing groups.

Turning to the analysis of hate speech detection, we find that the average percentage of tweets containing hate speech is $14.9 \pm 0.2$ in Andalusia and $13.1 \pm 0.3$ in Barcelona. Both in Andalusia and Barcelona, polarization and hate speech are anti-correlated before the election date, while this correlation is not statistically significant afterward. Notably, in both cases, the anti-correlation disappears after detrending, indicating that the fluctuations in polarization and hate speech are not directly correlated.

In Andalusia, the anti-correlation between polarization and hate speech before the election can be explained by the increasing participation of users from the block opposing Vox. As more users from this opposing block engage in the discussion, overall polarization rises, but the prominence of hate speech tweets decreases, since these users tend not to employ hate speech in their discourse. Additionally, \cite{capdevila2022emergencia} have suggested that far-right parties like Vox may moderate their speech during electoral campaigns due to the pressure of Spanish electoral law.

A key limitation of this study is the difference in scope between regional and city council elections. While regional elections often address broader issues, municipal elections focus on local concerns. Nevertheless, city council elections can still generate significant attention, especially in cities with complex political landscape.

In summary, our findings underscore the significant role that populist far-right actors play in shaping polarization patterns. While far-right parties tend to polarize discourse around two ideological blocks, other dichotomous actors, such as independence movement, do not, because of other transversal political stances diffusing the polarization. In both cases, polarizing actors are the responsible for most of the hate speech present and the observed trends suggest consistent anti-correlation between hate speech and polarization and hate speech and participation. 

\bibliography{references}

\section*{Funding}
This work was conducted as part of the Project PLEC2021-007850 funded by MCIN/AEI /10.13039/501100011033 and by the European Union NextGenerationEU/PRTR. E.R is a fellow of Eurecat’s “Vicente López” PhD grant program, while J.M. was a fellow of the same program. E.C. acknowledges support from the Spanish grants PGC2018-094754-B-C22 and PID2021-128005NB-C22, funded by MCIN/AEL 10.13039/501100011033 and “ERDF A way of making Europe", and from Generalitat de Catalunya under project 2021-SGR-00856.

\section*{Author contributions}
J.V and E.C conceptualised the study, J.M developed the methodology, performed data analyses, E.R performed data analyses, wrote the original draft and all authors
critically discussed the results, revised the paper and approved the final manuscript.

\section*{Competing interests}
The authors declare that they have no competing interests. 

\section*{Data availability}
The data that support the findings of this study are available from the corresponding author upon request. Additionally, the tweet IDs can be accessed in our GitHub repository at \url{https://github.com/remiss-project/remiss-data}.

\newpage

\renewcommand{\thefigure}{S1.\arabic{figure}} 
\setcounter{figure}{0} 

{
\centering
\section*{Supplementary Material}

\subsection*{Far-right party influence on polarisation dynamics in electoral campaigns}

Eva Rifà, Joan Massachs, Emanuele Cozzo, Julian Vicens

}

\subsubsection*{S1. Extended Stability of Communities}

This section provides additional results for the Section 3.4 ``Stability of communities" from the main article. Like in the main article, we compute the probability of a user to change from one community at one specific day to another community the day after. 

\begin{figure}[h!]
\centering
\includegraphics[width=0.8\textwidth]{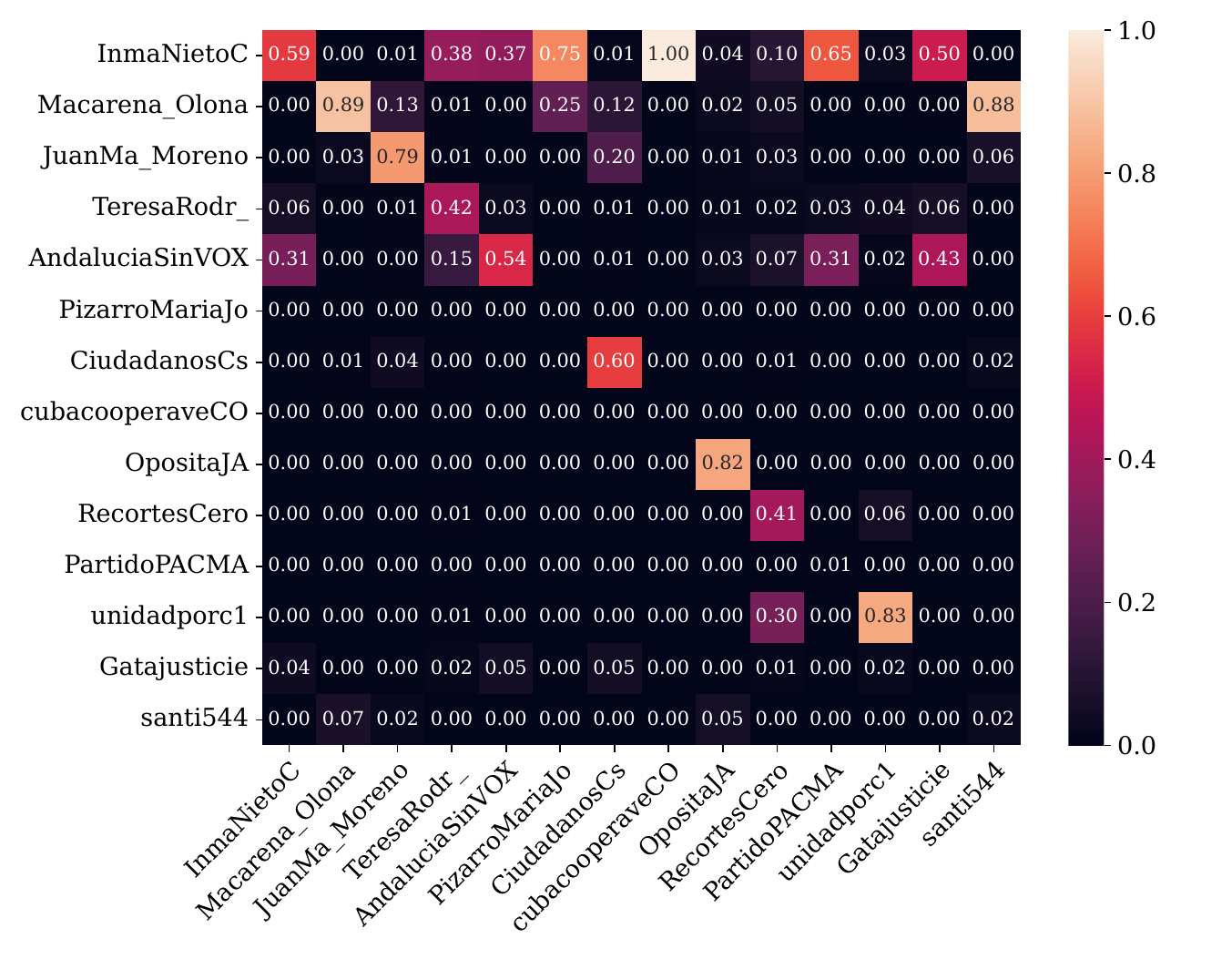}
\caption{Estimation of probability of a user belonging to community X in a day to become a member of the community Y the day after for Andalusia dataset. Communities are labelled with their most central user.}
\label{prob_and2_extended}
\end{figure}

\begin{figure}[h]
\centering
\includegraphics[width=0.8\textwidth]{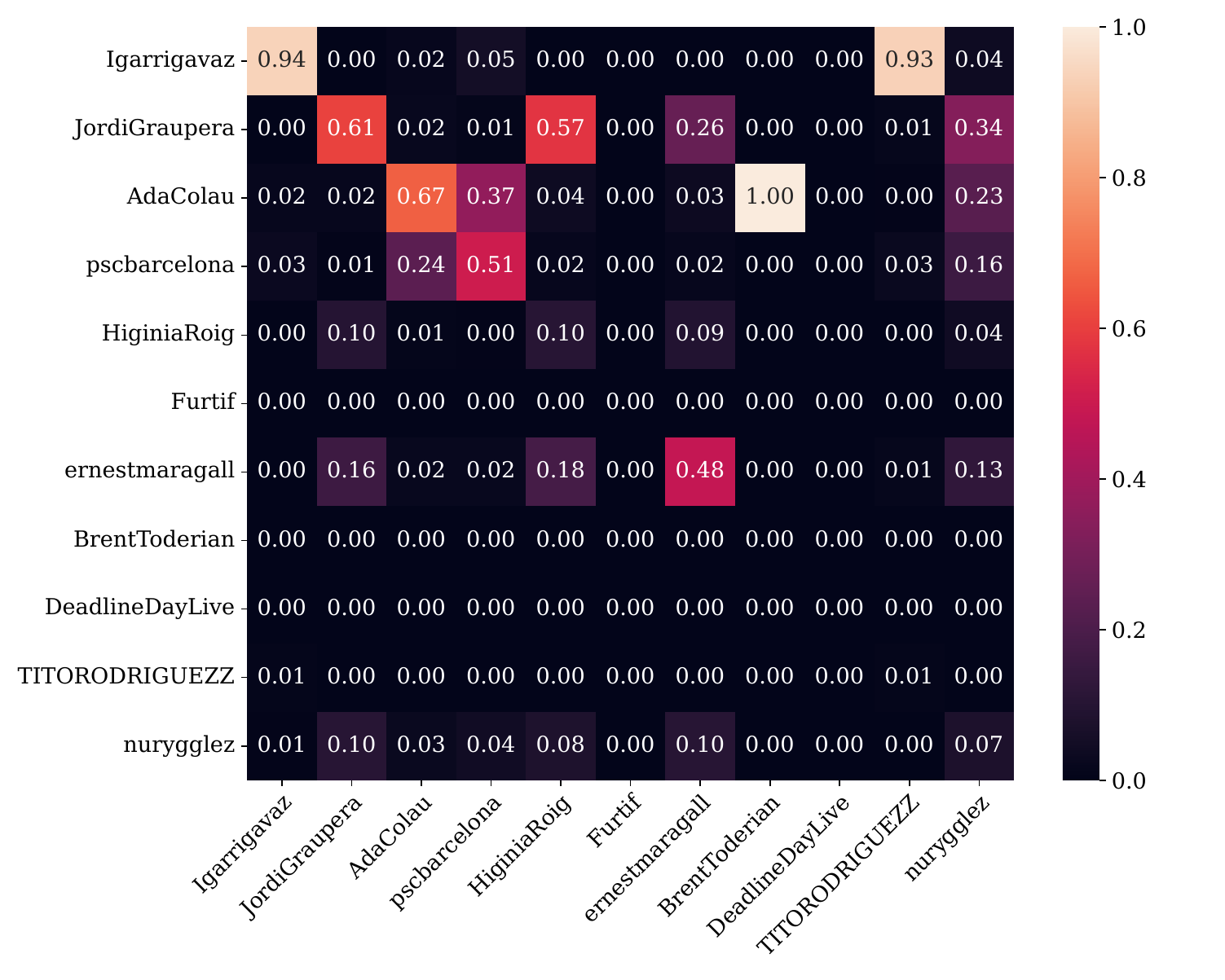}
\caption{Estimation of probability of a user belonging to community X in a day to become a member of the community Y the day after for Barcelona dataset. Communities are labelled with their most central user.}
\label{prob_bcn19_extended}
\end{figure}

By comparison, Fig. 10 in the main article contains only the communities that are labelled, the ones associated with the main political parties participating in the electoral race. However, here additional communities found by the algorithm are included. These smaller communities are described in Table 1 for Andalusia and in Table 2 for Barcelona. Fig. \ref{prob_and2_extended} shows the community change probabilities for Andalusia dataset while Fig. \ref{prob_bcn19_extended} presents the corresponding data for Barcelona. For both cases each community is represented by the username of the central node. Also, they are ordered by the number of users.

In Andalusia, we observe that the community centered around \textit{@ImmaNietoC} (PorA) has the highest number of other communities with non-null probabilities of sending users, indicating greater connectivity compared to other communities. Interestingly, the most stable community is associated with \textit{@Macarena$\_$Olona} (Vox), while those with the lowest probabilities of retaining users are linked to the following usernames: \textit{@PizarroMariaJo}, \textit{@cubacooperaveCO}, \textit{@Gatajusticie}, \textit{@PartidoPACMA} and \textit{@santi544}.

In Barcelona, \textit{@JordiGraupera} (Junts community) and \textit{AdaColau} (BComú) are the central nodes in the communities that attract the most users from other groups. The most stable community is led by \textit{@Igarrigavaz} (Vox), which also has a 0.9 probability of receiving users exclusively from the community centered on \textit{@TITORODRIGUEZZ}. Conversely, the least stable labelled community is associated with \textit{@HiginiaRoig} (CUP community). Lastly, it is worth noting that there are two communities detected by the algorithm that have zero probability everywhere.

\renewcommand{\thefigure}{S2.\arabic{figure}} 
\setcounter{figure}{0} 

\subsubsection*{S2. Voter Perceptions of Political Ideology in Andalusia}

\begin{figure}[h]
\centering
\includegraphics[width=\textwidth]{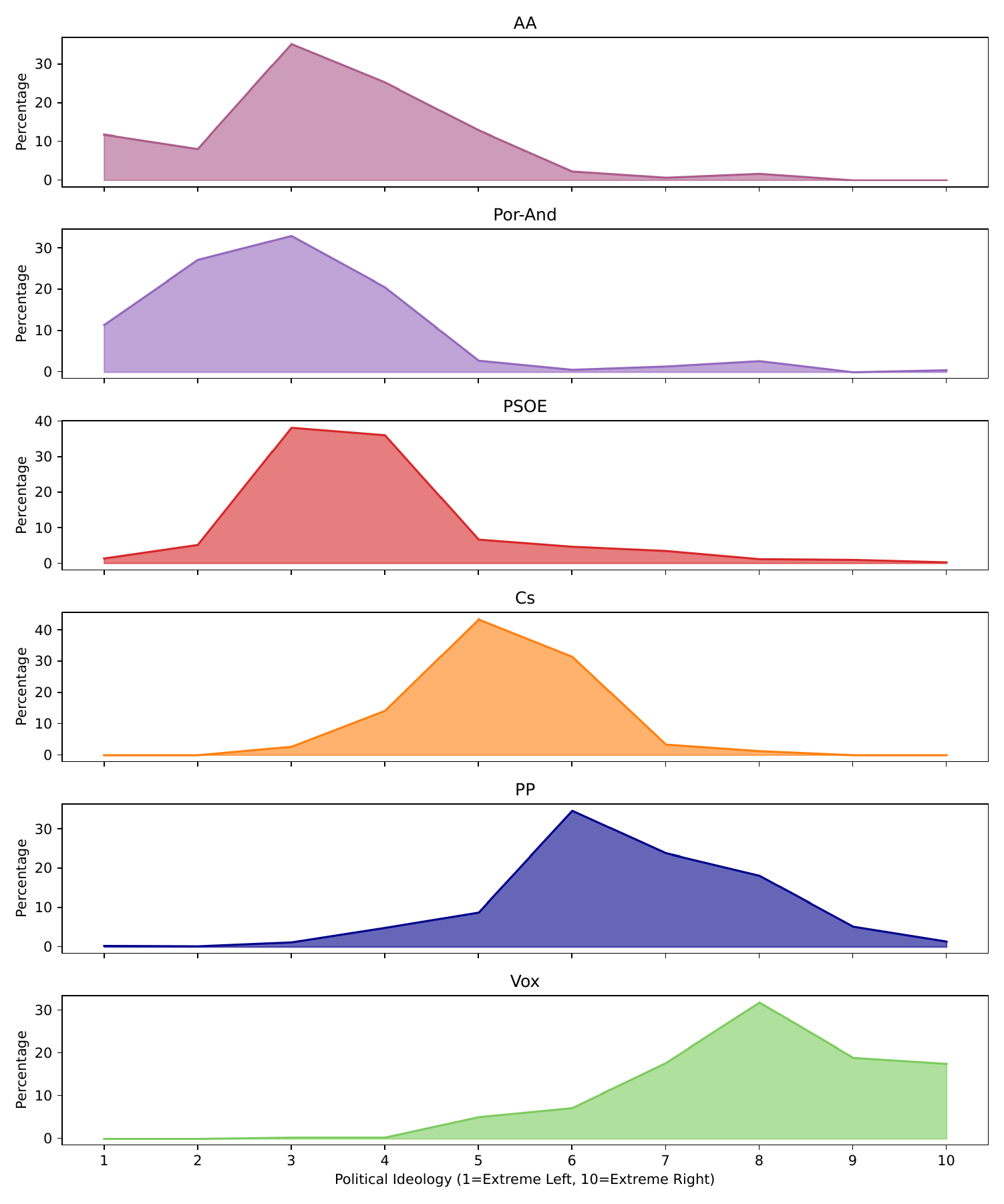}
\caption{Perception of political ideology for each Andalusian party based on the ``\textit{Barómetro Andaluz, Estudio de Opinión Pública de Andalucía (March 2022)}"}
\label{ideology-and}
\end{figure}

The \textit{Centro de Estudios Andaluces} conducted interviews to gather the perceptions of Andalusian society on politics, electoral behaviour, and other topics (e.g., the Ukraine-Russia conflict, the economy, etc.). In the ``\textit{Barómetro Andaluz, Estudio de Opinión Pública de Andalucía}" ~\citep{CEA2022}, with data collected in March 2022, citizens were asked specific questions regarding how they perceived the parties they voted for. One question asked voters of a particular party: \textit{``When talking about politics, the terms 'left' and 'right' are often used. On a scale of 1 to 10, where 1 is 'far left' and 10 is 'far right,' where would you place yourself?"}. This information provides insight into voters' perceptions of Andalusian political parties (Fig. \ref{ideology-and}).

\renewcommand{\thefigure}{S3.\arabic{figure}} 
\setcounter{figure}{0} 

\subsubsection*{S3. Voter Perceptions of Political Ideology and Catalan Independence in Catalonia}

\begin{figure}[h]
\centering
\includegraphics[width=\textwidth]{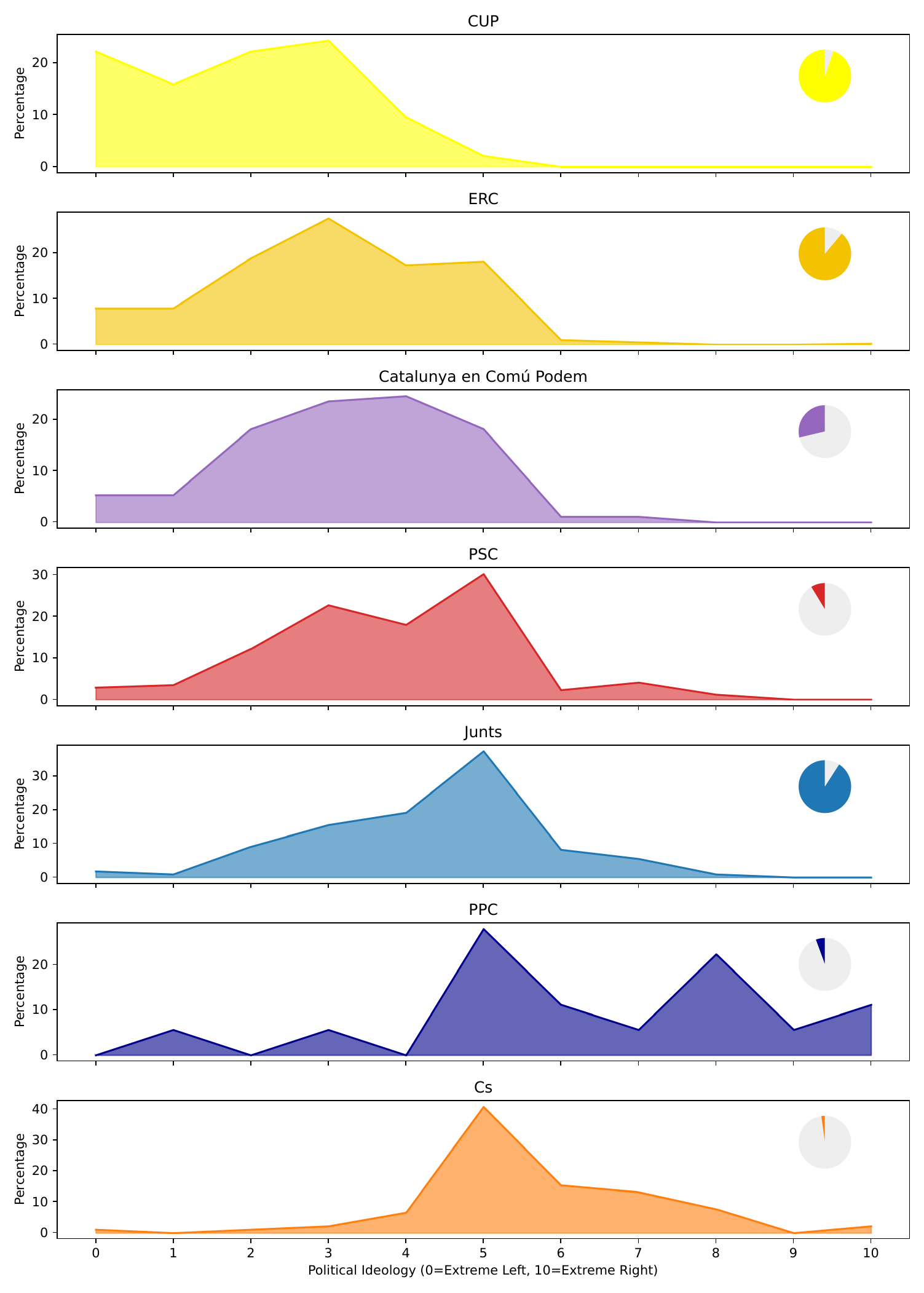}
\caption{Perception of political ideology and position of Catalan independence for each Catalan party based on the ``\textit{Baròmetre d'Opinió Política (March 2019)}"}
\label{ideology-cat}
\end{figure}

The CEO (Centre d’Estudis d’Opinió) conducted a personal survey to gather the perceptions of Catalan society on politics, media, electoral behaviour, and the evaluation of political leaders. In the first wave of ``Baròmetre d’Opinió Política" ~\citep{CEO2019}, with data collected in March 2019, citizens were asked two specific questions regarding how they perceived the parties they voted for. The first question asked voters of a particular party: ``\textit{When people talk about politics, they often use the terms 'left' and 'right'. Could you tell me where you would place yourself on a scale from 0 to 10, where 0 means far left and 10 means far right? (Parliament of Catalonia)}". The second question asked: ``\textit{Do you want Catalonia to become an independent state?}". This information provides insight into voters' perceptions of Catalan political parties, covering the ideological spectrum and their position on Catalan independence (Fig. \ref{ideology-cat}).

\end{document}